\documentclass[preprint]{aastex}

\shorttitle{HD~63021 is an Ae Star}
\shortauthors{Whelan et al}

\begin{document}

\title{HD~63021: An Ae Star with X-ray Flux}

\author{David G. Whelan}
\affil{Department of Physics, Austin College, 900 N. Grand Avenue, Sherman, TX 75090}
\email{dwhelan@austincollege.edu}

\author{Jon Labadie-Bartz}
\affil{Department of Physics, Lehigh University}

\author{S. Drew Chojnowski}
\affil{Department of Astronomy, New Mexico State University}

\author{James Daglen}
\affil{Daglen Observatory, Mayhill, NM}

\author{Ken Hudson}
\affil{Grey Tree Observatory, Mayhill, NM}


\section{Spectroscopic Variability}

Balmer and Fe~{\sc ii}~(42) multiplet emission were discovered in a
spectrum of HD~63021 on 10~April~(UTC), 2018. Subsequent observations
revealed variability in both photospheric absorption lines and Balmer
line emission.

\begin{figure*}[h]
\includegraphics[scale=0.3]{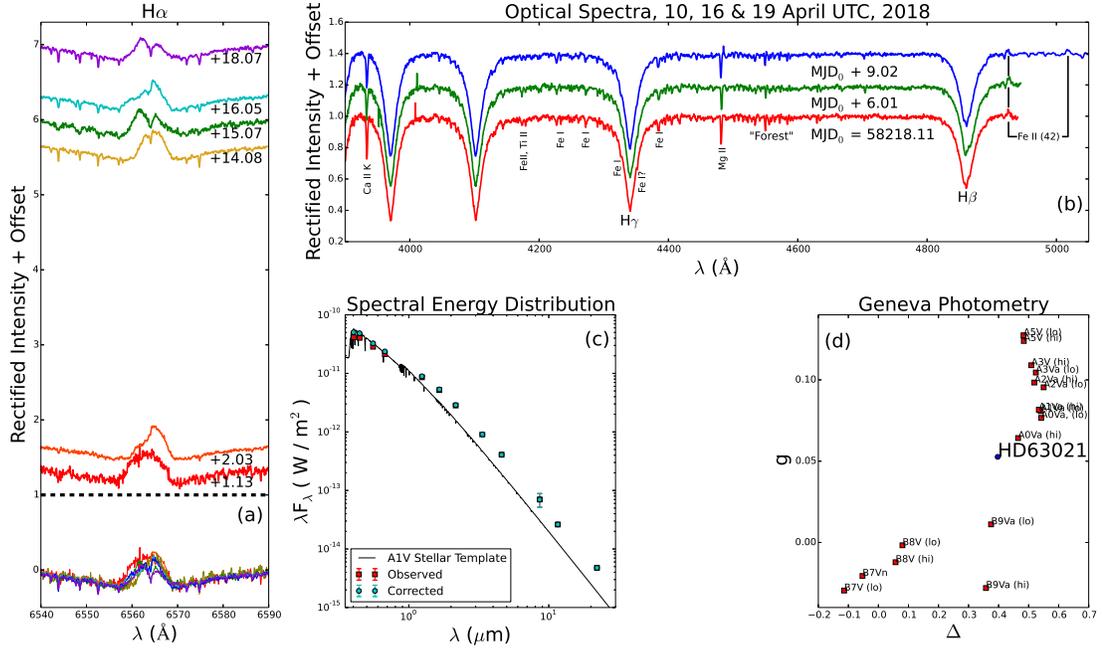}
\caption{(a) The H$\alpha$\ observations from Daglen and Grey Tree
  Observatories. On bottom all spectra are overplotted, on top they
  are offset by a factor of days from first optical spectrum. (b) The
  UV-optical spectra taken at Adams Observatory. (c) The spectral
  energy distribution, showing the IR excess. (d) Geneva
  photometry.\label{fig}}
\end{figure*}

Figure~\ref{fig}~(a) shows H$\alpha$\ observations over the course of
18~nights. The emission morphology changes on a nightly basis;
additional spectra not exhibited show that emission changes
perceptibly on the scale of hours.

Three optical spectra (Figure~\ref{fig}~(b)) demonstrate the spectral
type, A1V(e), and photospheric absorption line variability. We
determined the spectral type primarily from the Balmer line wing shape
and metal line absorption strengths. The absorption line strength
variability is on the $\sim20$ \% level for the Ca~{\sc ii}~K, Fe~{\sc
  i}, Fe~{\sc ii}/Ti~{\sc ii}, and Mg~{\sc ii}\ lines labeled. We do
not detect such variability in the Si~{\sc ii}~$\lambda$~4128-4130 and
Ca~{\sc i}~$\lambda$~4226\ lines, among others. The presence of such
pronounced Fe~{\sc i} lines is unusual for A1V stars, as they are
usually indicative of slightly later spectral types ({\em e.g.,}
A3-5V).

\section{Archival Data}

We plot the available infrared photometry using 2MASS, WISE, and AKARI
data in Figure~\ref{fig}~(c) to show that there is a substantial
infrared excess beginning in the near-infrared that roughly follows
the stellar Rayleigh-Jeans tail.

Photometric surveys have been used to estimate fundamental parameters
for this star, with varying results. The \citet{McDonald12} and
\citet{Cotten16} surveys of stars with infrared excesses claim that
this is an $\sim$A6 star with a surface temperature of
$\sim$7300-7400K. Gaia~DR2 \citep{Gaia18} quotes a large
A$_G$\ ($\sim$0.43 magnitude) and a similarly low temperature. The
Tycho-2 photometry \citep{Wright03}, however, claims a surface
temperature of 10,500~K and a spectral type of B9. Since extinction
correction is an issue we pulled the Geneva photometry
\citep{Rufener88,Mermilliod97}, and plot the reddening-free parameters
g and $\Delta$\ in Figure~\ref{fig}~(d) against spectral standards of
low vsini (``lo'') and high vsini (``hi'') from
\citet{Gray87,Garrison94}. This plot shows that HD~63021 is located
between A0V and B9V stars. Since the observed spectral type is A1V,
this is a bit of a surprise.

Using the Gaia~DR2 parallax, the observed V-band magnitude from
\citet{Hog00}, and our derived A$_V$ of 0.146, we determine an
absolute magnitude of M$_V = 0.76$. This is very close to the B9.5V
absolute magnitude from \citet{Pecaut13}.

In addition, HD~63021 is an X-ray source. Its X-ray luminosity is
$1.02 \times 10^{31}$~erg/s, using XMM~Newton band~8 (0.2-2.0~keV)
data \citep{XMM17} and the Gaia~DR2 parallax.

\section{Discussion and Future Work}

The H$\alpha$\ emission exhibits rapid V/R variability, a
near-constant emission line strength, and narrow peak separation. The
constant emission line strength suggests to us a star that is
continuously feeding its decretion disk, which in turn suggests a star
spinning close to its break-up speed. Since the absorption lines are
narrow and the H$\alpha$\ peak separation is small, it seems likely
that we are viewing this source at low inclination.

The H$\alpha$\ V/R variability is coupled with photospheric absorption
line strength changes, possibly from filling in by circumstellar disk
continuum emission. In addition, the strong excess in the near-IR
suggests free-free emission from gas that is quite hot
\citep{Reig11}. We wonder whether the Tycho-2 and Geneva photometry
suggests a higher surface temperature because of the hot circumstellar
gas, a binary companion, or both.

The X-ray luminosity is much higher than other A-type stars that emit
X-rays \citep[$\sim 10^{22-24}$~erg/s; ][]{Schroder07}. One
possibility is that the stellar wind is very strong, similar to that
of B-type stars \citep{Cassinelli94}. Another option is an X-ray
binary \citep{Reig11}. In the case of an X-ray binary, then there
would be a compact source associated with HD~63021.

We will be conducting an optical spectroscopic monitoring program to
look for signs of periodicity, analyses of lightcurves from
ground-based surveys and the space-based TESS mission, and will
additionally create a comparison sample of X-ray sources of similar
spectral type from available archives. New infrared observations would
also be beneficial, in order to make a contemporary SED.

\acknowledgements 

This work would not have been possible without the Adams Observatory
at Austin College. We have made use of data from the European
Space Agency (ESA) mission {\it Gaia}
(\url{https://www.cosmos.esa.int/gaia}), processed by the {\it Gaia}
Data Processing and Analysis Consortium (DPAC,
\url{https://www.cosmos.esa.int/web/gaia/dpac/consortium}).


\begin{thebibliography}{}

\bibitem[Cassinelli et al.(1994)]{Cassinelli94} Cassinelli, J.~P.,
  Cohen, D.~H., Macfarlane, J.~J., Sanders, W.~T., \& Welsh,
  B.~Y. 1994, \apj, 421, 705.

\bibitem[Cotten \& Song(2016)]{Cotten16} Cotten, T.~H., \& Song,
  I. 2016, \apjs, 225, 15.

\bibitem[Gaia Collaboration(2018)]{Gaia18} Gaia Collaboration, et
  al. 2018, \aap, submitted.

\bibitem[Garrison \& Gray(1994)]{Garrison94} Garrison, R.~F., \& Gray,
  R.~O. 1994, \aj, 107, 1556.

\bibitem[Gray \& Garrison(1987)]{Gray87} Gray, R.~O., \& Garrison,
  R.~F. 1987, \apjs, 65, 581.

\bibitem[H\o g et al.(2000)]{Hog00} H\o g, E., Fabricius, E., Makarov,
  V.~V., Urban, S., Corbin, T., Wycoff, G., Bastian, U., Schwekendiek,
  P., \& Wicenec, A. 2000, \aap, 355, 27.

\bibitem[McDonald, Zijlstra, \& Boyer(2012)]{McDonald12} McDonald, I.,
  Zijlstra, A.~A., \& Boyer, M.~L. 2012, \mnras, 427, 343.

\bibitem[Mermilliod  et  al.(1997)]{Mermilliod97}  Mermilliod,  J.-C.,
  Mermilliod, M., \& Hauck, B. 1997, \aaps, 124, 349.

\bibitem[Pecaut \& Mamajek(2013)]{Pecaut13} Pecaut, M.~J., \& Mamajek,
  E.~E. 2013, \apjs, 208, 9.

\bibitem[Reig(2011)]{Reig11} Reig, P. 2011, \apss, 332, 1.

\bibitem[Rufener \& Nicolet(1988)]{Rufener88} Rufener, F., \& Nicolet,
  B. 1988, \aap, 206, 357.

\bibitem[Schr\"{o}der \& Schmitt(2007)]{Schroder07} Schr\"{o}der, C.,
  \& Schmitt, J.~H.~M.~M. 2007, \aap, 475, 677.

\bibitem[Wright et al.(2003)]{Wright03} Wright, C.~O., Egan, M.~P.,
  Kraemer, K.~E., \& Price, S.~D. 2003, \aj, 125, 359.

\bibitem[XMM-SSC(2017)]{XMM17} XMM-Newton selw survey Source
  Catalogue, version 2.0 (XMM-SSC, 2017). Leicester, UK, 2017,
  2018yCat.9053....0X.

\end{thebibliography}
\end{document}